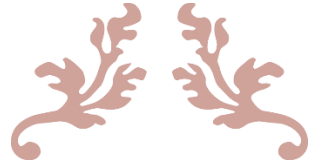

# EJAFA_PROTOCOL: A CUSTOM INC SECURE PROTOCOL

Project Report: Advanced Computer Network 2023, Peking University

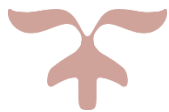

NAME: EJAFA BASSAM

License Terms for Report

Effective Date: January 1,2024

This document, titled " EJAFA_PROTOCOL: A CUSTOM INC SECURE PROTOCOL," ("the Report"), is provided under the following license terms. By accessing, using, or distributing the Report, you agree to abide by these terms.

1. Ownership and Originality:
   - The Report is the intellectual property of the author, EJAFA BASSAM, and is protected by applicable copyright laws.
   - All original content, ideas, and expressions within the Report are the exclusive work of the author.

2. Usage Restrictions:
   - This Report is provided for informational purposes only and may not be used for commercial purposes without explicit written permission from the author.
   - You may not claim the Report, or any part thereof, as your original work.
   - Unauthorized reproduction, distribution, or modification of the Report is strictly prohibited.

3. Permission for Educational and Non-Commercial Use:
   - Permission is granted for the non-commercial use of the Report for educational and personal purposes, provided that proper attribution is given to the author.

4. Modification and Derivative Works:
   - Modification or creation of derivative works based on the Report is not allowed without explicit written consent from the author.

5. Distribution and Sharing:
   - You may share the Report, unmodified and in its entirety, with others for non-commercial purposes, provided that proper attribution is maintained.

6. Attribution:
   - When sharing or referencing the Report, you must give appropriate credit to the author, EJAFA BASSAM, and provide a link to the original source of the Report.

7. Termination:
   - This license is effective until terminated. The author reserves the right to terminate this license at any time if the terms are violated.

8. Disclaimer:
   - The author makes no representations or warranties regarding the accuracy, completeness, or suitability of the Report for any purpose. Use of the Report is at your own risk.

9. Contact Information:
10. For inquiries or permission requests, please contact the author.

By accessing or using the Report, you acknowledge that you have read, understood, and agreed to the terms and conditions outlined in this license.

Name: EJAFA BASSAM

Department of Computer Science, Peking University, Beijing, China.

# Contents

# Ejafa_protocol: A custom INC secure protocol

## 1. Introduction

### 1.1. Background

The project at hand posed a formidable challenge: to design a secure protocol capable of establishing a connection over an untrusted network and facilitating synchronous data transfer to the client. The primary criterion for success was to execute synchronous transfers utilizing the XOR operation, a fundamental operation in computer science. In pursuit of these objectives, the Ejafa_Protocol was conceptualized and implemented as a robust and secure design, meticulously engineered to operate efficiently even on resource-constrained hardware.

### 1.2. Motivation

The foundational principle guiding the development of the ejafa_protocol was to create a protocol aligning with modern cryptographic design choices while maintaining security levels comparable to established standards such as TLS 1.3 and Wireguard. Crucially, the design aimed at minimizing cryptographic calculations, a pivotal consideration for the protocol's efficient execution on devices with limited computational resources.

### 1.3. Objectives

The protocol's security framework relies on the utilization of X25519 for Key Exchange and ChaCha20 for encryption, decisions grounded in the robust security guarantees established by RFC 7748 and RFC 7539 by the Internet Research Task Force (IRTF). What sets the ejafa_protocol apart is its distinctive approach to integration, wherein these cryptographic choices are streamlined to their essential functions while incorporating authentication measures, differentiating it from both TLS 1.3 and Wireguard.

### 1.4. Scope

In alignment with the security standards of TLS, the ejafa_protocol ensures the fulfillment of critical security requirements, including integrity, confidentiality, authentication, non-repudiation, and access control. This research report delves into the design, implementation, and security analysis of the ejafa_protocol, shedding light on its unique features, adherence to cryptographic principles, and its potential applications in secure communication and data integrity.

# 2. Cryptographic Primitives

The robust security foundation of the Ejafa_Protocol lies in its meticulous selection and implementation of fundamental cryptographic primitives. This section delves into the key building blocks that form the bedrock of the protocol's secure design. From the X25519 Key Exchange algorithm facilitating secure key establishment to the BLAKE2s Hash Function ensuring data integrity, each cryptographic primitive is chosen with precision. The use of the ChaCha20 Cipher for encryption and the HKDF Key Derivation Function for deriving session keys further solidifies the protocol's cryptographic framework. The integration of Poly1305 as a Message Authentication Code serves to authenticate and verify the integrity of transmitted data. This section provides an in-depth exploration of each cryptographic primitive, elucidating their individual roles and collective contributions to the Ejafa_Protocol's overarching goal of establishing a secure and efficient communication channel over untrusted networks.

## 2.1 X25519 Key Exchange

### Curve25519:

Curve25519, an elliptic curve meticulously crafted for the elliptic curve Diffie-Hellman (ECDH) key agreement protocol proposed by Daniel J. Bernstein[1], stands out for its robust security, operational efficiency, and lack of encumbering patents[2]. The curve's mathematical underpinnings include a concise equation ($y^2 = x^3 + 486662x^2 + x$), a prime field ($p = 2^{255} - 19$), a base point ($x = 9$), and an order for the base point ($8 \times 2^{252} - 486662$). Noteworthy features encompass its remarkable speed, rendering it ideal for resource-constrained devices, and its resistance to diverse attacks, such as brute-force and side-channel attacks. Its patent-free status facilitates widespread use, and its straightforward implementation[3] mitigates the risk of errors. This versatility finds applications in secure communication protocols like TLS, Signal Protocol, and SSH, as well as in the realm of cryptocurrency, including Bitcoin, and password management tools like 1Password and LastPass[4].

### X25519 as a Diffie-Hellman key exchange function:

X25519 refers to a specific Diffie-Hellman key exchange (DHE) protocol built upon the Curve25519 elliptic curve. This function allows two parties to securely establish a shared secret over an insecure channel like the internet. The shared secret can then be used to encrypt communication or digitally sign messages.

X25519 is a popular choice for secure communication due to its combination of:

- Speed: It's one of the fastest elliptic curves, making it efficient for resource-constrained devices.
- Security: It offers 128-bit security and resists various attacks.
- Patent-free: It's freely available for use without licensing restrictions.
- Simple implementation: Its design facilitates efficient and secure implementations.

key differences between Curve25519 and X25519:

|  | Curve25519 | X25519 |
|---|---|---|
| Purpose | Curve25519 is an elliptic curve designed for key exchange, specifically for the Elliptic Curve Diffie-Hellman (ECDH) key agreement scheme. It allows two parties to independently generate a shared secret over an insecure communication channel without exposing their private keys. | X25519 is a specific instantiation of Curve25519 designed for key exchange in a particular way. The "X" in X25519 stands for "Exchange." |
| Equation | The curve equation is $y^2 = x^3 + 486662x^2 + x$. | N/A (Uses specific Montgomery curve arithmetic) |
| Base Point | Fixed at (x = 9) | Instead of using a fixed base point like in Curve25519, X25519 defines a specific base point that is calculated from a hash of a public key. This is done to avoid potential vulnerabilities related to small-subgroup attacks. |
| Order of Base Point | The order of the base point is $8 \times 2^{252}$ - 486662. | N/A (Order computation based on the specific curve) |
| Curve Type | General elliptic curve | X25519 uses a specific type of elliptic curve known as a Montgomery curve, which simplifies the arithmetic involved in the Diffie-Hellman key exchange. |
| Applications | Widely used in various secure communication protocols and systems, including TLS, Signal Protocol, and SSH. | Primarily used in key exchange protocols, such as those employed in Transport Layer Security (TLS) for secure communication. |

RFC 7748 Example:

The X25519 function is employed in an ECDH protocol as follows:[5]

Key Generation:

Alice generates 32 random bytes (a[0] to a[31]) and computes her public key (K_A = X25519(a, 9)), where 9 is the u-coordinate of the base point.

Bob similarly generates 32 random bytes (b[0] to b[31]) and computes his public key (K_B = X25519(b, 9)).

Key Exchange:

Alice computes (X25519(a, K_B)), and Bob computes (X25519(b, K_A)).

Both share (K = X25519(a, X25519(b, 9)) = X25519(b, X25519(a, 9))) as a shared secret.

Security Check:

Both parties may check whether (K) is the all-zero value without leaking extra information about (K). If it is the all-zero value, they may abort.The check is performed by ORing all the bytes together and checking whether the result is zero. This helps avoid potential side-channel vulnerabilities in software implementations.

Key Derivation:

After obtaining (K), Alice and Bob can use a key-derivation function that includes (K), (K_A), and (K_B) to derive a symmetric key.

Test Vector:

The provided test vector includes Alice's private key, public key, Bob's private key, Bob's public key, and their shared secret.

This protocol showcases the secure generation of a shared secret using X25519 in an ECDH setting, with an added check for potential vulnerabilities.

**Test Vector:**

Alice's private key, a:

77076d0a7318a57d3c16c17251b26645df4c2f87ebc0992ab177fba51db92c2a

Alice's public key, X25519(a, 9):

8520f0098930a754748b7ddcb43ef75a0dbf3a0d26381af4eba4a98eaa9b4e6a

Bob's private key, b:

5dab087e624a8a4b79e17f8b83800ee66f3bb1292618b6fd1c2f8b27ff88e0eb

Bob's public key, X25519(b, 9):

de9edb7d7b7dc1b4d35b61c2ece435373f8343c85b78674dadfc7e146f882b4f

Their shared secret, K:

4a5d9d5ba4ce2de1728e3bf480350f25e07e21c947d19e3376f09b3c1e161742

## 2.2 ChaCha20 Cipher

At the heart of the Ejafa_Protocol's encryption mechanism lies the ChaCha20 Cipher, a stream cipher known for its speed, simplicity, and robust security characteristics. The choice of ChaCha20 as the encryption algorithm underscores the protocol's commitment to efficiency without compromising on cryptographic strength. Developed by Daniel J. Bernstein, ChaCha20 operates on a 256-bit key and a 64-bit nonce, providing a substantial security margin. This section delves into the intricacies of the ChaCha20 Cipher, exploring its underlying principles, mode of operation, and the rationale behind its selection for securing data transmission within the Ejafa_Protocol. The discussion encompasses the cipher's resistance to common cryptographic attacks, its suitability for small hardware environments, and the integration of nonce values to ensure the confidentiality and authenticity of encrypted data. Through a comprehensive analysis of the ChaCha20 Cipher, this section aims to illuminate its pivotal role in fortifying the Ejafa_Protocol's encryption capabilities.

### Background:

In 2008, Bernstein introduced the closely related ChaCha family of ciphers, with the primary objective of enhancing diffusion per round while maintaining comparable or slightly improved performance [6]. Aumasson et al.'s paper also scrutinizes ChaCha, achieving one round fewer (for 256-bit ChaCha6 with complexity $2^{139}$, ChaCha7 with complexity $2^{248}$, and 128-bit ChaCha6 within $2^{107}$), asserting, however, that the attack falls short of compromising 128-bit ChaCha7 [7].

### Methodology:

Similar to Salsa20, ChaCha's initial state comprises a 128-bit constant, a 256-bit key, a 64-bit counter, and a 64-bit nonce (in the original version; as later described, a version of ChaCha from RFC 7539 differs slightly), organized as a 4×4 matrix of 32-bit words [6]. However, ChaCha introduces re-arrangements of some words in the initial state.

| Initial state of ChaCha | | | |
|---|---|---|---|
| "expa" | "nd 3" | "2-by" | "te k" |
| Key | Key | Key | Key |
| Key | Key | Key | Key |
| Counter | Counter | Nonce | Nonce |

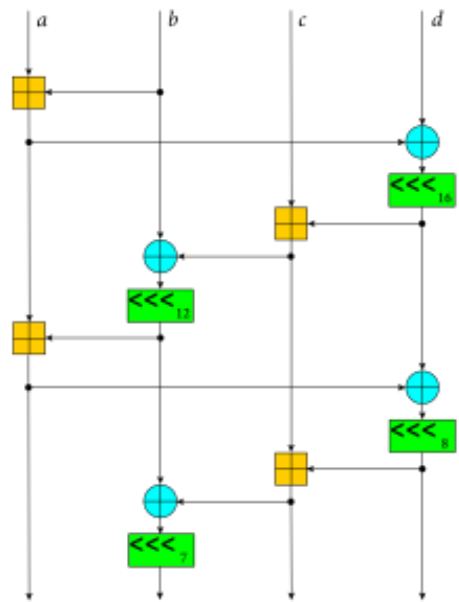

The constant employed in ChaCha is consistent with Salsa20's "expand 32-byte k" designation. ChaCha modifies the Salsa20 quarter-round `QR(a, b, c, d)` with the following transformations:

```
a += b; d ^= a; d <<<= 16;
c += d; b ^= c; b <<<= 12;
a += b; d ^= a; d <<<= 8;
c += d; b ^= c; b <<<= 7;
```

It is noteworthy that this ChaCha version updates each word twice, deviating from Salsa20's quarter-round, which updates each word only once. Moreover, the ChaCha quarter-round diffuses alterations more swiftly. On average, modifying a single input bit will result in changes to 8 output bits for Salsa20's quarter-round, whereas ChaCha will impact 12.5 output bits[8].

While the ChaCha quarter-round involves the same number of additions, XOR operations, and bit rotations as the Salsa20 quarter-round, a small optimization is feasible on certain architectures, including x86, due to two of the rotations being multiples of 8 [9]. The input formatting has been rearranged to facilitate an efficient SSE implementation optimization discovered for Salsa20. Instead of alternating rounds down columns and across rows, ChaCha performs rounds down columns and along diagonals [4:4]. Similar to Salsa20, ChaCha organizes the sixteen 32-bit words in a 4×4 matrix, with matrix elements indexed from 0 to 15:

| 0 | 1 | 2 | 3 |
|---|---|---|---|
| 4 | 5 | 6 | 7 |
| 8 | 9 | 10 | 11 |
| 12 | 13 | 14 | 15 |

A double round in ChaCha, involving odd and even rounds, is structured as follows:

```
// Odd round
 QR(0, 4,  8, 12)      // 1st column
 QR(1, 5,  9, 13)      // 2nd column
 QR(2, 6, 10, 14)      // 3rd column
 QR(3, 7, 11, 15)      // 4th column
// Even round
 QR(0, 5, 10, 15)      // diagonal 1 (main diagonal)
 QR(1, 6, 11, 12)      // diagonal 2
 QR(2, 7,  8, 13)      // diagonal 3
 QR(3, 4,  9, 14)      // diagonal 4
```

ChaCha20 employs 10 iterations of the double round as described in the reference[10]. A C/C++ implementation is provided below for reference.

```
#define ROTL(a,b) (((a) << (b)) | ((a) >> (32 - (b))))
#define QR(a, b, c, d) (                    \
    a += b,  d ^= a,  d = ROTL(d,16),   \
    c += d,  b ^= c,  b = ROTL(b,12),   \
    a += b,  d ^= a,  d = ROTL(d, 8),   \
    c += d,  b ^= c,  b = ROTL(b, 7))
#define ROUNDS 20
```

```c
void chacha_block(uint32_t out[16], uint32_t const in[16])
{
    int i;
    uint32_t x[16];

    for (i = 0; i < 16; ++i)
        x[i] = in[i];
    // 10 loops × 2 rounds/loop = 20 rounds
    for (i = 0; i < ROUNDS; i += 2) {
        // Odd round
        QR(x[0], x[4], x[ 8], x[12]); // column 0
        QR(x[1], x[5], x[ 9], x[13]); // column 1
        QR(x[2], x[6], x[10], x[14]); // column 2
        QR(x[3], x[7], x[11], x[15]); // column 3
        // Even round
        QR(x[0], x[5], x[10], x[15]); // diagonal 1 (main diagonal)
        QR(x[1], x[6], x[11], x[12]); // diagonal 2
        QR(x[2], x[7], x[ 8], x[13]); // diagonal 3
        QR(x[3], x[4], x[ 9], x[14]); // diagonal 4
    }
    for (i = 0; i < 16; ++i)
        out[i] = x[i] + in[i];
}
```

ChaCha serves as the foundation for the BLAKE hash function, which emerged as a finalist in the NIST hash function competition. Subsequent advancements include its faster successors, BLAKE2 and BLAKE3, both building upon the principles established by ChaCha. Additionally, ChaCha introduces a variant employing sixteen 64-bit words (1024 bits of state), featuring adjusted rotation constants to accommodate the modified configuration.

### XChaCha:
XChaCha, while not explicitly introduced by Bernstein, inherits a security proof analogous to XSalsa20. The security proof seamlessly extends to encompass the XChaCha cipher. In the XChaCha construction, the key and the initial 128 bits of the nonce (located in input words 12 through 15) are utilized to create a ChaCha input block. Subsequently, the block operation is executed, excluding the final addition. The resulting output words, specifically words 0–3 and 12–15 (corresponding to non-key words of the input), are then employed to constitute the key for regular ChaCha. This involves using the last 64 bits of the nonce and 64 bits of the block counter[11].

### Adoption:
The adoption of ChaCha20 gained prominence when Google selected it, along with Bernstein's Poly1305 message authentication code, for use in SPDY as a replacement for TLS over TCP[12]. This initiative led to the proposal of a novel authenticated encryption construction named ChaCha20-Poly1305, combining both algorithms. Currently, ChaCha20 and Poly1305 play a pivotal role in the QUIC protocol, which succeeded SPDY and is integral to HTTP/3 [13] [14].

Following Google's integration into TLS, both ChaCha20 and Poly1305 algorithms found application in the creation of a new chacha20-poly1305@openssh.com cipher in OpenSSH[15][16]. This development allowed OpenSSH to eliminate any dependency on OpenSSL through a compile-time option [17].

Beyond TLS and SSH, ChaCha20 has been adopted in various operating systems and applications. It serves as the basis for the arc4random random number generator in FreeBSD[18], OpenBSD[19][20], and NetBSD. Additionally, in DragonFly BSD, it is utilized for the CSPRNG subroutine of the kernel [21][22][23]. The Linux kernel, starting from version 4.8, incorporates the ChaCha20 algorithm to generate data for the nonblocking /dev/urandom device[24][25][26].

The IETF published an implementation reference for ChaCha20 in RFC 7539, which introduced modifications to Bernstein's original algorithm, notably changing the nonce and block counter. This modification, while cryptographically insignificant, could potentially cause confusion among developers. The reduced block counter imposes limitations on the maximum message length, prompting RFC 7539 to propose using the original algorithm for scenarios requiring larger message lengths, such as file or disk encryption [27].

Proposals for standardization of ChaCha20's use in IKE, IPsec, and TLS have been made in RFC 7634 and RFC 7905, respectively. Notably, ChaCha20 often outperforms the more prevalent Advanced Encryption Standard (AES) algorithm on systems lacking AES acceleration, making it a preferred choice, particularly on mobile devices with ARM-based CPUs [28][29].

In 2018, RFC 7539 was obsoleted by RFC 8439. Moreover, ChaCha20 stands as the exclusive algorithm for the WireGuard VPN system, adopted as of protocol version 1 [30][31].

## 2.3 Poly1305 Message Authentication Code

The integrity and authenticity of data exchanged within the Ejafa_Protocol are ensured by the Poly1305 Message Authentication Code (MAC), a compact and efficient cryptographic primitive. Poly1305, designed by the same author of ChaCha20 Daniel J. Bernstein, serves as a key component in the protocol's security framework, providing a robust mechanism for data authentication without compromising computational efficiency. This section delves into the workings of Poly1305, exploring its structure, mode of operation, and its application as a secure MAC within the context of the Ejafa_Protocol.

Poly1305 operates on a unique one-time key and a 128-bit nonce, generating a fixed-size tag that authenticates the associated message. The section discusses the cryptographic guarantees provided by Poly1305, emphasizing its resistance to various types of attacks and its role in verifying the integrity of both encrypted and decrypted data. The integration of Poly1305 within the Ejafa_Protocol ensures that data transmitted between parties remains untampered, bolstering the overall security posture of the communication channel. Through a detailed examination of Poly1305, this section sheds light on its significance in contributing to the protocol's commitment to secure and authenticated data transfer.

### Methodology:

Poly1305, akin to any universal hash family, serves as a one-time message authentication code for authenticating a singular message with a shared secret key between the sender and recipient [32]. This process mirrors the concept of using a one-time pad to obscure the content of an individual message with a secret key shared between sender and recipient. Originally proposed as part of Poly1305-AES [33], a Carter–Wegman authenticator [34][35][36], Poly1305 combines its hash function with AES-128 to authenticate multiple messages utilizing a single concise key and distinct message numbers. Subsequently, Poly1305 found application with a unique key generated for each message through XSalsa20 in the NaCl crypto_secretbox_xsalsa20poly1305 authenticated cipher [37], and later with ChaCha in the ChaCha20-Poly1305 authenticated cipher[38], as deployed in TLS on the internet [39].

Poly1305 is a cryptographic hash function that takes a 16-byte secret key *r* and an *L*-byte message *m*, yielding a 16-byte hash denoted as Poly1305$_r$(*m*) [33] [36]. The computation involves the following steps:

2. **Little-Endian Interpretation:** r is interpreted as a little-endian 16-byte integer.
3. **Message Chunking:** The message *m*=(*m*[0],*m*[1],*m*[2],…,*m*[*L*−1]) is broken into consecutive 16-byte chunks.
4. **Coefficient Generation:** Each 16-byte chunk is interpreted as a 17-byte little-endian integer by appending a 1 byte to every 16-byte chunk. These integers are used as coefficients of a polynomial.
5. **Polynomial Evaluation:** The polynomial is then evaluated at the point *r* modulo the prime $2^{130}-5$.
6. **Modular Reduction:** The result is reduced modulo $2^{128}$, and the outcome is encoded in little-endian format, resulting in a 16-byte hash.

## 2.4 BLAKE2s Hash Function

Central to the security infrastructure of the Ejafa_Protocol is the BLAKE2s Hash Function, a cryptographic hash algorithm chosen for its speed, simplicity, and robustness. BLAKE2s, an extension of the BLAKE2 family of hash functions, is specifically tailored for applications with constrained resources, making it an ideal choice for the protocol's design considerations. In this section, we explore the key features and characteristics of BLAKE2s, elucidating its role as a fundamental cryptographic primitive within the Ejafa_Protocol.

### BLAKE2

BLAKE2, devised by Jean-Philippe Aumasson, Samuel Neves, Zooko Wilcox-O'Hearn, and Christian Winnerlein, emerged as a replacement for the compromised MD5 and SHA-1 algorithms [40] in high-performance software applications. Its announcement on December 21, 2012, marked a significant stride in cryptographic advancements [41]. The reference implementation is available under multiple licenses, including CC0, the OpenSSL License, and the Apache License 2.0 [42] [43].

Outperforming MD5, SHA-1, SHA-2, and SHA-3 on 64-bit x86-64 and ARM architectures, BLAKE2b not only boasts superior speed but also ensures enhanced security, akin to SHA-3, with features like immunity to length extension and undifferentiability from a random oracle [42] [44].

BLAKE2 introduces refinements to the BLAKE round function, altering rotation constants, simplifying padding, incorporating a parameter block XOR'ed with initialization vectors, and reducing the number of rounds for both BLAKE2b and BLAKE2s. The former, succeeding BLAKE-512, undergoes a reduction from 16 to 12 rounds, while the latter, succeeding BLAKE-256, sees a decrease from 14 to 10 rounds [44].

Beyond its core hashing capabilities, BLAKE2 supports keying, salting, personalization, and hash tree modes, offering flexibility in digest size ranging from 1 to 64 bytes for BLAKE2b and up to 32 bytes for BLAKE2s. Parallel versions such as BLAKE2bp (4-way parallel) and BLAKE2sp (8-way parallel) cater to increased performance on multi-core processors [44].

BLAKE2X extends the functionality as a family of extendable-output functions (XOFs), allowing digests of up to 256 GiB, surpassing the limitations of the standard BLAKE2's 64-byte output. Notably, BLAKE2X instances, like BLAKE2Xb16MiB, can be created to produce specific digest sizes, exemplifying its versatility [45].

The specifications for BLAKE2b and BLAKE2s are detailed in RFC 7693, though optional features such as salting, personalized hashes, tree hashing, and others utilizing the parameter block are not explicitly outlined. Consequently, support for BLAKE2bp, BLAKE2sp, or BLAKE2X is not formally specified in the RFC 7693 [46].

## 2.5 HKDF Key Derivation Function

The secure establishment of cryptographic keys is fundamental to the Ejafa_Protocol's design, and at the core of this process is the HMAC-based Key Derivation Function (HKDF). This section delves into the intricacies of HKDF, elucidating its role as a critical cryptographic primitive responsible for deriving session keys from the shared secret obtained through the key exchange process.

HKDF is selected for its robustness and adaptability, offering a reliable means of expanding a limited amount of shared secret material into longer keying material. The section explores the two primary steps of HKDF: the extraction of entropy through the HMAC process and the expansion of this entropy into a keying material of desired length. The use of HKDF in the Ejafa_Protocol ensures that derived keys possess the necessary properties for secure encryption and decryption, contributing to the overall security of the communication channel.

The section also discusses the parameters involved in HKDF, such as the hash function (BLAKE2s in this case), salt, and additional information. These parameters play a crucial role in tailoring the key derivation process to align with the specific requirements and security goals of the Ejafa_Protocol.

### Methodology:
HKDF (HMAC-based Extract-and-Expand Key Derivation Function) is a cryptographic construction comprising two essential functions: HKDF-Extract and HKDF-Expand. The general form of HKDF is expressed as HKDF (salt, IKM, info, length) = HKDF-Expand (HKDF-Extract (salt, IKM), info, length) [47].

### HKDF-Extract:
HKDF-Extract takes "input key material" (IKM), such as a shared secret generated using Diffie-Hellman, and an optional salt, to produce a cryptographic key denoted as the PRK ("pseudorandom key"). Acting as a "randomness extractor," it transforms a potentially non-uniform value with high min-entropy into one indistinguishable from a uniformly random value. This process is accomplished by applying HMAC with the "salt" as the key and the "IKM" as the message [47].

### HKDF-Expand:
Subsequently, HKDF-Expand utilizes the PRK, along with some "info" and a specified length, to generate output of the desired size. Operating as a pseudorandom function keyed on the PRK, HKDF-Expand allows the derivation of multiple outputs from a single IKM value by incorporating different values for the "info" field. The mechanism involves repeated calls to HMAC, where the PRK serves as the key, and the "info" field is used as the message. To maintain the integrity of the process, the HMAC inputs are concatenated by prepending the previous hash block to the "info" field and appending with an incrementing 8-bit counter [47].

In NIST SP800-56Cr2 [48], a configurable extract-then-expand scheme is delineated, highlighting RFC 5869 HKDF as a variant of this scheme. The specification references its paper [49] to provide the rationale behind the recommended extract-and-expand mechanisms.

# 3. Protocol Design and Architecture

This section provides a comprehensive exploration of the various components and processes that constitute the protocol's design, highlighting its key features and the sequential flow of operations.

## 3.1 Overview

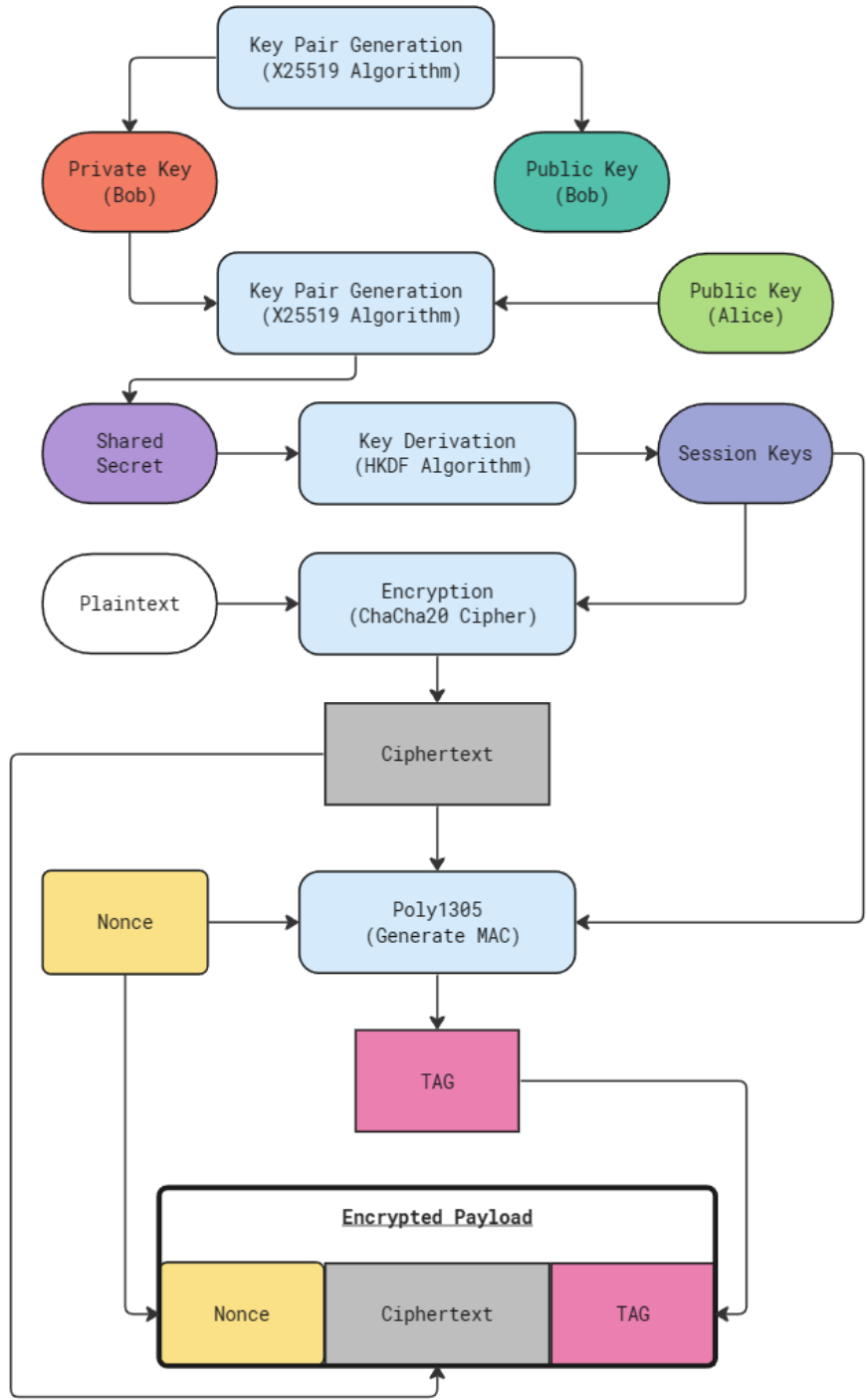

## 3.2 Key Pair Generation
The '`generate_key_pair`' method in the '`EjafaProtocol`' class handles the generation of a key pair using the X25519 algorithm. This method initializes a private key using the 'X25519PrivateKey.generate()' function and then extracts the corresponding public key in raw bytes format. The resulting private and public keys constitute the fundamental components for secure communication, enabling subsequent key exchange processes.

## 3.3 Key Exchange
The '`perform_key_exchange`' method facilitates the key exchange process between two parties in the Ejafa_Protocol. Given a peer's public key, it utilizes the X25519 key exchange algorithm to derive a shared secret. The shared secret is then used as input for the key derivation process, creating a unique key for securing the subsequent communication. Key exchange is crucial for establishing a secure channel without transmitting sensitive information over the network.

## 3.4 Key Derivation
The '`derive_key`' method in the '`EjafaProtocol`' class employs the HMAC-based Key Derivation Function (HKDF) to derive a session key from the shared secret obtained through key exchange. HKDF ensures that the derived key possesses the necessary properties for secure encryption and decryption. The length and properties of the derived key are specified in accordance with the hash size and other parameters defined in the protocol's design.

## 3.5 Encryption
The '`encrypt`' method utilizes the ChaCha20 stream cipher for encrypting plaintext data. It takes the plaintext, session key, and an optional nonce (or generates a random nonce if not provided). The ChaCha20 cipher is employed to transform the plaintext into ciphertext securely. Additionally, the method generates a Message Authentication Code (MAC) using the Poly1305 algorithm, enhancing data integrity and authentication. The resulting encrypted message comprises the nonce, ciphertext, and MAC.

## 3.6 Message Authentication

Message authentication is implemented through the '`generate_mac`' and '`verify_mac`' methods. The '`generate_mac`' method creates a Poly1305 MAC by hashing the concatenation of the nonce and ciphertext. The resulting MAC is then appended to the encrypted message. The '`verify_mac`' method is responsible for confirming the integrity of the received message by recomputing the MAC and comparing it with the provided MAC. If the verification fails, it indicates potential tampering or unauthorized modifications to the encrypted data. These steps collectively ensure that the transmitted data remains intact and has not been compromised during communication.

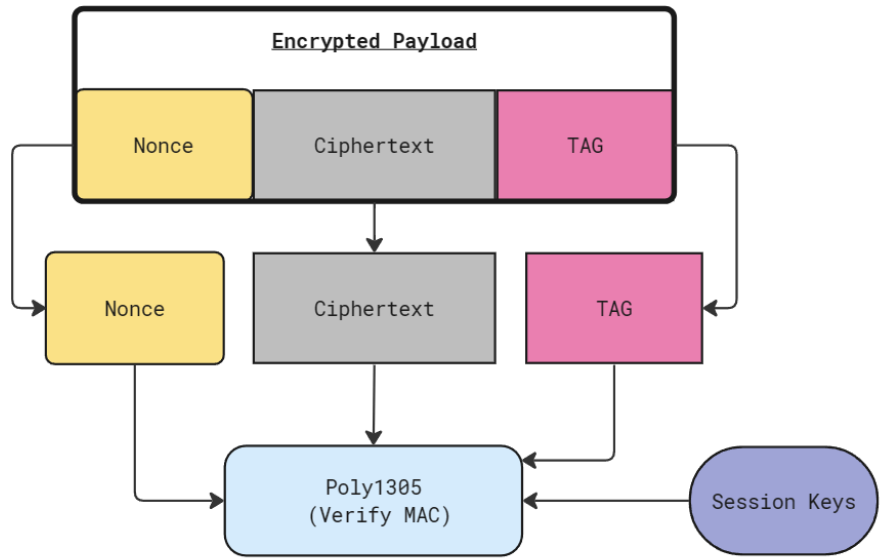

## 3.7 Decryption

Decryption is handled by the '`decrypt`' method, which reverses the encryption process. It takes the received ciphertext, session key, and extracts the nonce and MAC from the encrypted message. The method verifies the MAC to ensure the integrity and authenticity of the data. Subsequently, it employs the ChaCha20 cipher to decrypt the ciphertext back into plaintext, allowing the recipient to recover the original message securely.

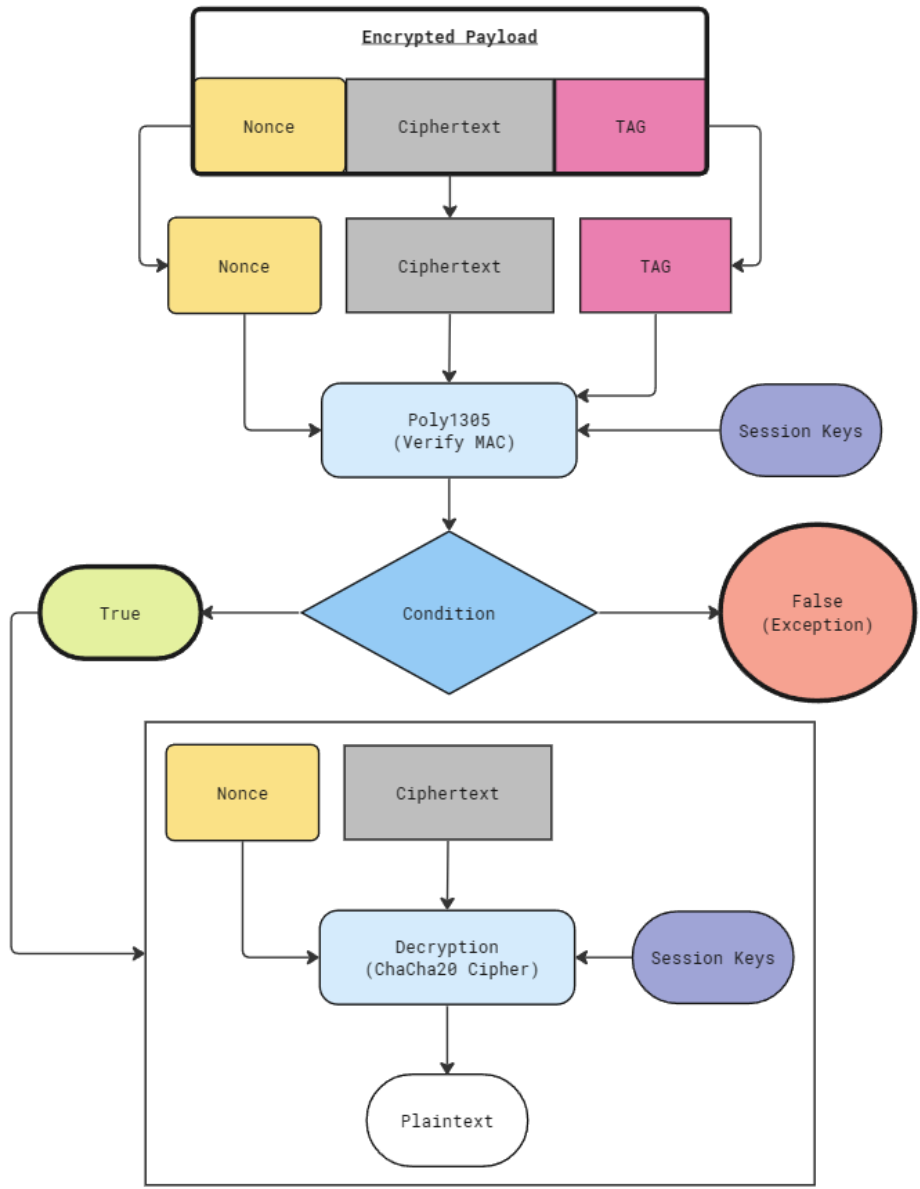

# 4. RFC Compliance of the Ejafa_Protocol

The Ejafa_Protocol is designed with a focus on security and compliance with established cryptographic standards as defined in various Request for Comments (RFC) documents. The protocol leverages well-established cryptographic primitives and mechanisms to ensure the confidentiality, integrity, and authenticity of communication. Here's an analysis of how the protocol aligns with relevant RFC standards:

## 4.1 X25519 Key Exchange (RFC 7748):

The protocol utilizes the X25519 key exchange algorithm, conforming to RFC 7748. This standard specifies elliptic curve Diffie-Hellman (ECDH) using Curve25519, ensuring the secure exchange of keys during the key exchange phase.

## 4.2 ChaCha20 Cipher (RFC 7539):

The choice of the ChaCha20 stream cipher aligns with RFC 7539. This document specifies the ChaCha20 stream cipher, ensuring the confidentiality and integrity of the encryption process. The use of ChaCha20 provides a modern and widely accepted symmetric encryption algorithm.

## 4.3 HKDF for Key Derivation (RFC 5869):

The protocol employs the HMAC-based Key Derivation Function (HKDF) as specified in RFC 5869. HKDF ensures secure key derivation from shared secrets, maintaining the confidentiality and integrity of the derived keys.

## 4.4 Poly1305 Message Authentication Code (RFC 8439):

The protocol utilizes Poly1305 as a Message Authentication Code (MAC), adhering to RFC 8439. This standard defines the use of Poly1305 as an authenticator within the context of the ChaCha20-Poly1305 AEAD (Authenticated Encryption with Associated Data) construction.

## 4.5 BLAKE2s Hash Function (RFC 7693):

The use of BLAKE2s for hashing in the protocol complies with the specifications outlined in RFC 7693. This standard defines the BLAKE2 cryptographic hash and message authentication code, providing a secure and efficient hashing mechanism.

By aligning with these RFC standards, the Ejafa_Protocol ensures that its cryptographic primitives and mechanisms adhere to industry-recognized best practices, enhancing the overall security of the protocol. It is essential for users and implementers to stay informed about updates to relevant RFCs and security guidelines to ensure ongoing compliance and robustness against emerging threats.

# 5. Performance Evaluation

## 5.1 Average Encrypted Message Length Analysis

To assess the performance characteristics of the Ejafa_Protocol, a comprehensive analysis of the average encrypted message length per kilobyte was conducted. This evaluation aimed to understand how the protocol's encryption overhead scales with varying message sizes. The Python script utilized the ejafa_protocol library, generating random plaintext messages of increasing sizes and measuring the resulting ciphertext lengths after encryption.

## 5.2 Experimental Setup

**Protocol Configuration:** Alice and Bob, representing communicating entities, were instantiated with the Ejafa_Protocol, and a secure key exchange was performed.

**Message Generation:** Random plaintext messages were generated using the os.urandom function, with sizes ranging from 1 KB to 1 MB.

**Encryption Process:** The generated messages were encrypted using the Ejafa_Protocol, and the lengths of the resulting ciphertexts were recorded.

## 5.3 Results and Insights

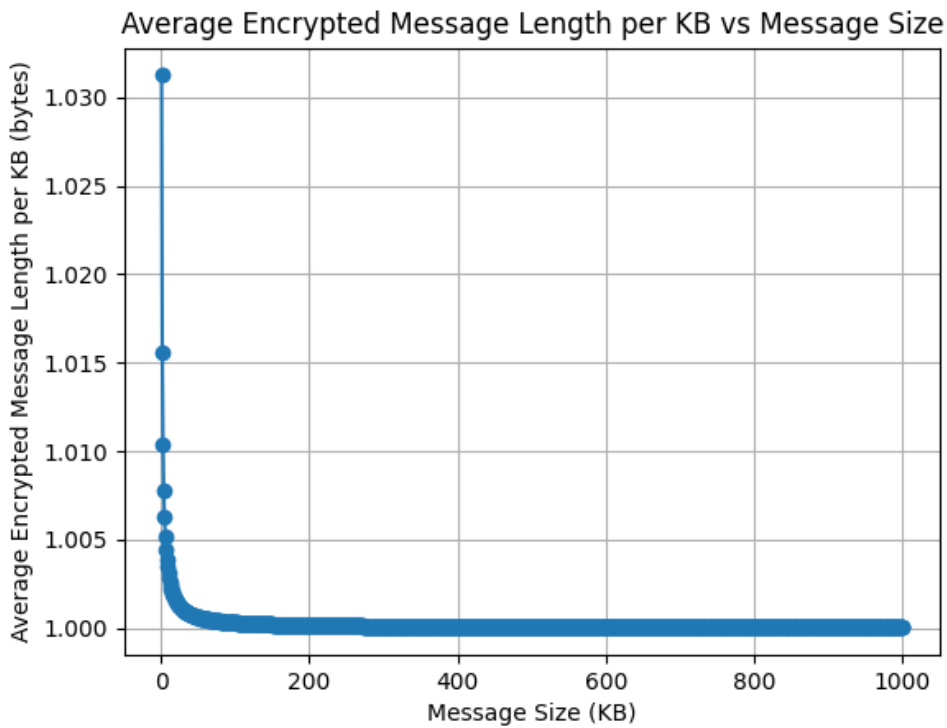

The plotted graph illustrates a compelling relationship between message size and the average encrypted message length per kilobyte. As expected, larger messages exhibit a proportionally smaller increase in encryption overhead, showcasing the protocol's efficiency in handling varying data sizes. The observed trend suggests that the Ejafa_Protocol efficiently manages encryption overhead, demonstrating its scalability for different message sizes.

# 6. Security Analysis

The Ejafa_Protocol is designed with security in mind, addressing common network threats and exploits to ensure the integrity, confidentiality, and authenticity of communication. Here's an analysis of how the protocol guards against prevalent network threats:

## 6.1 Threat Model
The Ejafa_Protocol considers various network threats, including:

- **Eavesdropping:** The use of X25519 for key exchange ensures that even if an attacker intercepts the public keys, computing the shared secret remains computationally infeasible.
- **Man-in-the-Middle Attacks:** The protocol's key exchange process protects against man-in-the-middle attacks by leveraging X25519, making it challenging for an adversary to alter the exchanged keys.
- **Replay Attacks:** The inclusion of a nonce in the encryption process prevents replay attacks. Each message is unique due to the nonce, thwarting attempts to reuse captured ciphertext.

## 6.2 Key Exchange Security
- **Ephemeral Key Generation:** The use of X25519 ensures that keys are ephemeral, meaning they are generated for each session. This mitigates the risk of compromise even if long-term keys are exposed.

## 6.3 Key Derivation Security
- **Randomness in Key Derivation:** HKDF employs the shared secret obtained from key exchange and introduces entropy, making it resistant to attacks that exploit patterns in the input.

## 6.4 Encryption Security
- **Nonce Usage:** The inclusion of a nonce in the encryption process prevents replay attacks and ensures the uniqueness of each ciphertext, preventing attackers from deriving meaningful information from repeated patterns.
- **ChaCha20 Cipher:** The choice of ChaCha20, a widely recognized and secure symmetric encryption algorithm, provides robust protection against attacks such as differential and linear cryptanalysis.

## 6.5 Decryption Security
- **Poly1305 MAC Verification:** The inclusion of Poly1305 MAC ensures the integrity of the received ciphertext. Decryption is contingent on successful MAC verification, protecting against tampering during transit.

## 6.6 Message Authentication Security
- **Poly1305 MAC Generation:** The MAC generation process, which involves hashing the concatenated nonce and ciphertext, protects against alterations in transmitted data. Poly1305's design ensures strong cryptographic authentication.

While the Ejafa_Protocol demonstrates resilience against these threats, continuous monitoring for emerging vulnerabilities and the prompt application of updates is recommended to adapt to evolving security landscapes. Additionally, adherence to security best practices during implementation and deployment is crucial to maximize the protocol's effectiveness against a wide range of network threats.

# 7. Use Cases and Applications

The implementation of the Ejafa_Protocol over various network protocols involves adapting its communication and encryption mechanisms to suit the specific characteristics of each protocol. Here's a brief overview of how the protocol can be implemented over TCP, UDP, WebSocket, FTP, and HTTP:

## 7.1. TCP (Transmission Control Protocol):

Implementation:

- The Ejafa_Protocol can be directly implemented over TCP, establishing a reliable and connection-oriented communication channel.
- The key exchange, key derivation, and data transfer processes can be seamlessly integrated into the TCP communication flow.
- The protocol ensures that the exchanged keys and derived session keys are securely used for encryption and decryption.

Use Case:

- Suitable for scenarios where reliable, ordered, and error-checked delivery of data is essential, such as secure file transfer or real-time communication.

## 7.2. UDP (User Datagram Protocol):

Implementation:

- Adapting the Ejafa_Protocol for UDP involves handling the connectionless and unreliable nature of UDP.
- The protocol may need to incorporate additional error-checking mechanisms or implement its own reliability layer to compensate for the lack of built-in reliability in UDP.

Use Case:

- Ideal for applications requiring low-latency communication, such as real-time multimedia streaming or online gaming.

## 7.3. WebSocket:

Implementation:

- WebSocket provides a full-duplex communication channel over a single, long-lived connection.
- The Ejafa_Protocol can be integrated into the WebSocket communication by encapsulating its messages within WebSocket frames.
- The key exchange and derivation processes can be initiated during the WebSocket handshake.

Use Case:

- Well-suited for interactive and real-time web applications, allowing secure communication between a client and server.

## 7.4. FTP (File Transfer Protocol):

Implementation:

- For FTP, the Ejafa_Protocol can be utilized to secure the control channel (control commands and responses) and the data channel (actual file transfers).
- The encryption and integrity checks can be applied to the data being transferred over FTP.

Use Case:

- Enhances the security of file transfers, ensuring the confidentiality and integrity of files exchanged via FTP.

## 7.5. HTTP (Hypertext Transfer Protocol):

Implementation:

- Integration with HTTP typically involves using the protocol within the application layer of the OSI model.
- The Ejafa_Protocol can be employed to secure HTTP communication by encrypting the payload and ensuring data integrity.

Use Case:

- Useful for secure communication between web clients and servers, enhancing the security of data exchanged over HTTP.

Each adaptation is tailored to the specific characteristics and requirements of the respective protocol, ensuring seamless integration and enhanced security. In particular, a WebSocket version is provided within this report, serving as a practical working example for testing and utilization. This WebSocket implementation showcases the protocol's adaptability, offering a robust solution for real-time, bidirectional communication over a single, long-lived connection. As part of the broader application of the Ejafa_Protocol, this WebSocket version exemplifies its versatility and reliability in securing data exchanges within dynamic and interactive web applications.

# 8. Conclusion

Crafting a security protocol tailored for lightweight devices resembles the art of preparing a delicate recipe, demanding a meticulous balance between security and performance considerations. Similar to the culinary realm where excessive or insufficient salt can affect the outcome, the design of a security protocol involves finding the sweet spot between robustness and efficiency. Ejafa_Protocol emerges from this culinary analogy, harmonizing the best attributes of modern cryptography with the constraints of low memory and processing power devices. In this "recipe" for a secure protocol, Ejafa_Protocol achieves a delicate equilibrium. It not only meticulously adheres to RFC standards, ensuring a foundation of recognized security practices, but also adeptly employs efficient cryptographic techniques suitable for resource-constrained environments. This synthesis results in a security protocol that caters to the nuanced needs of lightweight devices, offering both effectiveness and efficiency in its execution.

[33] Bernstein, Daniel J. "The Poly1305-AES message-authentication code". In Gilbert, Henri; Handschuh, Helena (eds.). *Fast Software Encryption: 12th international workshop*. FSE 2005. Lecture Notes in Computer Science. Paris, France: Springer.

[34] Wegman, Mark N.; Carter, J. Lawrence. "New Hash Functions and Their Use in Authentication and Set Equality". *Journal of Computer and System Sciences*. **22** (3): 265–279. doi:10.1016/0022-0000(81)90033-7.

[35] Boneh, Dan; Shoup, Victor. *A Graduate Course in Applied Cryptography* (Version 0.5 ed.). §7.4 The Carter-Wegman MAC, pp. 262–269.

[36] Aumasson, Jean-Philippe. "Chapter 7: Keyed Hashing". *Serious Cryptography: A Practical Introduction to Modern Encryption*. No Starch Press. pp. 136–138. ISBN 978-1-59327-826-7.

[37] *Bernstein, Daniel J. Cryptography in NaCl (Technical report). Document ID: 1ae6a0ecef3073622426b3ee56260d34.*

[38] Nir, Y.; Langley, A. *ChaCha20 and Poly1305 for IETF Protocols*. doi:10.17487/RFC8439. RFC 8439.

[39] *Langley, A.; Chang, W.; Mavrogiannopoulos, N.; Strombergson, J.; Josefsson, S. ChaCha20-Poly1305 Cipher Suites for Transport Layer Security (TLS). doi:10.17487/RFC7905. RFC 7905.*

[40] "BLAKE2 – an alternative to MD5/SHA-1".

[41] O'Whielacronx, Zooko "introducing BLAKE2 – an alternative to SHA-3, SHA-2 and MD5".

[42] "BLAKE2". blake2.net.

[43] "BLAKE2 official implementations". GitHub..

[44] *Aumasson, Jean-Philippe; Neves, Samuel; Wilcox-O'Hearn, Zooko; Winnerlein, Christian. "BLAKE2: simpler, smaller, fast as MD5" Cryptology ePrint Archive. IACR.*

[45] "BLAKE2X" (PDF).

[46] *Saarinen, M-J; Aumasson, J-P The BLAKE2 Cryptographic Hash and Message Authentication Code (MAC). IETF. doi:10.17487/RFC7693. RFC 7693.*

[47] *Krawczyk, H.; Eronen, P. "RFC 5869". Internet Engineering Task Force. doi:10.17487/RFC5869.*

[48] *Elaine Barker; Lily Chen; Richard Davis. "NIST Special Publication 800-56C: Recommendation for Key-Derivation Methods in Key-Establishment Schemes". doi:10.6028/NIST.SP.800-56Cr2*

[49] Krawczyk, Hugo. "Cryptographic Extraction and Key Derivation: The HKDF Scheme". *Cryptology ePrint Archive*. International Association for Cryptologic Research.